\documentclass[sigconf]{acmart}

\pdfoutput=1    
\usepackage{multirow}
\usepackage{color}
    
\copyrightyear{2021} 
\acmYear{2021} 
\setcopyright{acmcopyright}\acmConference[CHI '21]{CHI Conference on Human Factors in Computing Systems}{May 8--13, 2021}{Yokohama, Japan}
\acmBooktitle{CHI Conference on Human Factors in Computing Systems (CHI '21), May 8--13, 2021, Yokohama, Japan}
\acmPrice{15.00}
\acmDOI{10.1145/3411764.3445490}
\acmISBN{978-1-4503-8096-6/21/05}
\acmSubmissionID{1963}

\begin{document}

\title[PTeacher: a Pronunciation Training System]{{PTeacher: a Computer-Aided Personalized Pronunciation Training System with Exaggerated Audio-Visual Corrective Feedback}}

\makeatletter
\def\authornote#1{%
\g@addto@macro\@authornotes{%
      \stepcounter{footnote}\footnotetext{#1}}%
}
\makeatother

\author{Yaohua Bu$^{*}$}
\affiliation{Academy of Arts \& Design, Tsinghua University}

\author{Tianyi Ma$^{*}$}
\affiliation{Department of Computer Science and Technology, Tsinghua University}

\author{Weijun Li}
\affiliation{School of Information Science and Technology, Northeast Normal University}

\author{Hang Zhou}
\affiliation{The Chinese University of Hong Kong}

\author{Jia Jia$^{\dagger}$}
\affiliation{Department of Computer Science and Technology, Tsinghua University}

\author{Shengqi Chen}
\affiliation{Department of Computer Science and Technology, Tsinghua University}

\author{Kaiyuan Xu}
\affiliation{Department of Computer Science and Technology, Tsinghua University}

\author{Dachuan Shi}
\affiliation{Department of Computer Science and Technology, Tsinghua University}

\author{Haozhe Wu}
\affiliation{Department of Computer Science and Technology, Tsinghua University}

\author{Zhihan Yang}
\affiliation{Department of Computer Science and Technology, Tsinghua University}

\author{Kun Li}
\affiliation{SpeechX Limited}

\author{Zhiyong Wu}
\affiliation{Tsinghua-CUHK Joint Research Center for Media Sciences, Technologies and Systems, Shenzhen International Graduate School, Tsinghua University}

\author{Yuanchun Shi}
\affiliation{Department of Computer Science and Technology, Tsinghua University}

\author{Xiaobo Lu}
\affiliation{Academy of Arts \& Design, Tsinghua University}

\author{Ziwei Liu}
\affiliation{S-Lab, Nanyang Technological University}

\authornote{Equal contribution.}
\authornote{Corresponding author, jjia@tsinghua.edu.cn.}
\settopmatter{printacmref=false, printfolios=false}
\renewcommand{\shortauthors}{Yaohua Bu and Tianyi Ma, et al.}

\begin{abstract}
Second language (L2) English learners often find it difficult to improve their pronunciations due to the lack of expressive and personalized corrective feedback.
In this paper, we present Pronunciation Teacher~(\textit{PTeacher}), a Computer-Aided Pronunciation Training (CAPT) system that provides personalized exaggerated  audio-visual corrective feedback for mispronunciations. 
Though the effectiveness of exaggerated feedback has been demonstrated, it is still unclear how to define the appropriate degrees of exaggeration when interacting with individual learners.
To fill in this gap, we interview {100 L2 English learners and 22 professional native teachers} to understand their needs and experiences. 
Three critical metrics are proposed for both learners and teachers to identify the best exaggeration levels in both audio and visual modalities. %
Additionally, we incorporate the personalized dynamic feedback mechanism given the English proficiency of learners.
Based on the obtained insights, a comprehensive interactive pronunciation training course is designed to help L2 learners rectify mispronunciations in a more perceptible, understandable, and discriminative manner.
Extensive user studies demonstrate that our system significantly promotes the learners' learning efficiency.
\end{abstract}



\maketitle

\section{Introduction}
Pronunciation plays a crucial role in English learning for second-language (L2) learners. 
However, a majority of L2 English learners encounter difficulties in improving their pronunciation. 
On the one hand, they are prone to pronounce L2 words in the tongue of their first language,
which is called the negative influence of language transfer~\cite{strange1995speech,meng2007deriving,derwing2002esl,kartushina2014effects,cortes2005negative}. This causes multiple inconspicuous mispronunciations that can hardly be rectified by themselves.
On the other hand, professional native-speaking English teachers, who can diagnose pronunciation problems and give corrective feedback~\cite{lightbown2000they}, are in pressing demand. 
According to statistics provided by the British Council, while approximately 1.5 billion English learners~\cite{council2013english} exist, only 250 thousand native speakers are qualified to serve as English teachers. It is also reported~\cite{derwing2005second} that 67\% of ESL teachers survey in Canada have no training in pronunciation instruction.
The insufficiency of native English teachers is even worse in underdeveloped areas.

Driven by the demand, quite a number of Computer-Aided Pronunciation Training (CAPT) systems  have been proposed~\cite{badin2010visual,iribe2011generating,meng2012hierarchical,meng2014synthesizing,ning2015hmm,meng2010synthesizing} 
with feedback from both audio and visual modalities~\cite{lu2010castle,lu2012automatic,wong2010development,leung2013development,yuen2011enunciate}. While most of them focus on synthesizing natural speech, their systems are designed to show only the pairwise differences between the correctly pronounced phonemes and the mispronounced ones.
The teaching effect of such strategy proves to be less significant due to L2 learners' weaker perceiving ability~\cite{escudero2001role}. Particularly, the difficulty for perceiving and correcting pronunciation grows along with age~\cite{khul255b,polka1994developmental}. As a result, more \textit{perceptible} and \textit{distinctive} feedback is in demand.

In offline English classes, exaggerated feedback has been a effective feedback strategy for teachers to rectify the pronunciation of learners~\cite{mora2017relationship,polka1994developmental}. Alghamdi, et al.~\cite{alghamdi2017impact} have proven that visually exaggerated speech is capable of promoting the perceptual ability of learners. %
However, this direction has rarely been explored in the field of human-computer interactions. Only Zhao et al.~\cite{zhao2013audiovisual} propose a basic pipeline for audio-visual exaggeration, but focusing more on implementing than evaluation. No practical training paradigm is constructed. %
Consequently, the problem of what compromises a good exaggeration system remains unsettled. %

In this paper, we provide systematical studies towards designing personalized exaggerated feedback in a CAPT system named Pronunciation Teacher~(\textit{PTeacher}). We focus on constructing a participatory exaggerated pronunciation corrective system that emphasizes enlarging the teaching effectiveness in a \textit{user perception} point of view. Specifically, our system evaluates learners' pronunciation with real-time mispronunciation detection and diagnosis (\textit{MDD}) algorithm~\cite{li2016mispronunciation}. Then exaggerated feedback is given in the form of audio and articulated animation. Importantly, we point out that through \textit{extensive human evaluation}, the following key issues can be thoroughly discussed: \textbf{1)} how to define the fine-grained parameters of exaggeration for both audio and visual modalities so that corrective feedback can be distinctive enough; \textbf{2)} how 
personalized responses can be made to L2 learners with different degrees of proficiency; and \textbf{3)} how much the proposed designed training course can positively affect the learning efficiency of L2 learners compared with traditional CAPT system.

To this end, a total number of 100 L2 learners together with 22 professional native English teachers have participated in our studies.
Among them 30 learners and 22 teachers are responsible for determining the set of exaggeration parameters which renders the best feedback performances on three key aspects: \textit{distinguishability}, \textit{understandability} and \textit{perceptibility}. 
Afterward, we leverage comprehensive user studies on 30 learners to connect different levels of exaggeration feedback with different levels of proficiency, which has never been discussed before.
We group and evaluate the learners through both objective scores reported from \textit{MDD} and subjective evaluations from native teachers. 
Then the best personalized feedback level suitable for each learner can be determined.

Equipped with these necessary data and analysis, we include an interactive pronunciation training course into the \textit{PTeacher} system, aiming to improve learners' engagement. Along with the course, our system evaluates the English proficiency of each learner in a life-long manner and provides flexible exaggerated audio-visual corrective feedback. Therefore, personalized exaggeration can be given according to real-time mispronunciation detection from \textit{MDD} as well as accumulated evaluation. 
User studies on the system demonstrate that our \textit{PTeacher} with the exaggerated feedback enhances the learners' pronunciation accuracy by 14.19\% and 27.55\% for learners with a higher and lower degree of proficiency respectively, within a short time of learning (1 hour).

The contributions of our work are listed as follows:

\begin{itemize}
\item We define and identify the most suitable set of parameters in exaggerated corrective pronunciation feedback in both audio and visual modalities from three critical aspects.
\item  We propose personalized exaggerated feedback according to English proficiency of the learner.
\item We design the audio-visual corrective CAPT system, \textit{PTeacher}, which includes a pronunciation training course. The course can dynamically evaluate learners' English pronunciation proficiency in a life-long manner.

\item We support all of our findings and analysis with extensive user studies conducted on 100 second-language English learners and 22 professional native teachers. Comprehensive results demonstrate the advantage of our proposed exaggerated training system as well as the effectiveness of each module.
\end{itemize}

\section{Background}
\label{sec:2}
Computer-aided or assisted learning is an important research area in both human-computer interactions (HCI)~\cite{neri2008effectiveness,chalfoun2011subliminal} and language learning~\cite{golonka2014technologies,miller1990major}. We focus on design a Computer-Aided Pronunciation Training System (CAPT) with exaggerated audio-visual feedback.
In this section, we will first present the learning theories behind our audio-visual exaggerated feedback design (Section~\ref{sec:2.1}). Then we demonstrate the connections between our work and the recent advances in HCI (Section~\ref{sec:2.2}). Finally, we review and discuss the related works in the specific area of CAPT (Section~\ref{sec:2.3}).

\subsection{Learning Theories}
\label{sec:2.1}
\textbf{Theories on Audio-Visual Feedback.}
Information processing in speech and language communication is bi-modal. For example, language learners not only listen to the speaker but also observe the speaker's articulatory movements~\cite{king1985integration, chen1998audio}. In the visual modality, phoneticians have summarized that articulatory phonetics are strongly correlated with the manners and places of articulation~\cite{halle2000feature,titze1998principles,ogden2017introduction,crystal2011dictionary}, articulators~\cite{kelly2006teach} and airflow~\cite{geoghegan2012stereoscopic}. In the auditory modality, the principles of phonology and phonetics dictate and explain the ways humans make sounds~\cite{rogerson2011english,trubetzkoy1969principles,pierrehumbert1980phonology}. These theories lead to our design of providing both audio and visual information as feedback. While previous methods show improvements in L2 learners’ pronunciation abilities through training with articulatory animations~\cite{bliss2018computer,kartushina2015effect}, our user study shows that with exaggerated feedback from both modalities, the learning efficiency can be further improved compared to providing only exaggerated audio feedback, which supports the hypotheses of the theories.

\noindent\textbf{Exaggerated Feedback.}
Numerous studies have suggested that many L2 speech production (pronunciation) difficulties are rooted in perception~\cite{meng2007deriving, escudero2001role, escudero2005linguistic, ricard1986beyond, colantoni2015second}. Moreover, it has been exemplified that reinforcing the perception ability of learners can significantly contribute to the speech production ability automatically~\cite{bu2020visual,ortega1999enhancing, strange2008speech, lee2016effects, bradlow1997training, lee2020effects}. Exaggerated audiovisual feedback is a particular kind of perception reinforcement, which corrects the pronunciation by strengthening the user's visual or auditory attention. For example, by enhancing the duration of the audio of a nasal consonant, brain plasticity at the perceptual and pre-attentive neural levels can be strengthened~\cite{cheng2019temporal, cheng2019role}. Increasing the movement of the animation, the user's visual perception of graphics can be enhanced. Specifically, exaggerated movement can lead to more memorable perception than non-exaggerated movement~\cite{granqvist2018exaggeration}. Therefore, we propose the exaggerated-feedback to improve the perceptual effect of the target phonemes in both audio and visual modalities.

\subsection{Language Learning and Exaggerated Feedback in HCI}
\label{sec:2.2}
As there are rarely any studies that target exaggerated feedback in language learning specifically in HCI, we look into the studies that contribute mostly to the sub-area of language learning and exaggerated feedback.
Previous works on language learning mainly focus on the effectiveness of computer aided training and the components that matters. Ambra et al.~\cite{neri2008effectiveness} investigated whether a language learning system can help young L2 learners improve word-level pronunciation skills. They also provided a fundamental evaluation of the effectiveness of computer-assisted language learning systems. Robertson et al.~\cite{robertson2018designing}, advocated that new interactive designs supporting collaboration can be used to overcome engineering limitations. In our work, we also propose interactive courses in our system. Hailpern~\cite{hailpern2009creating} introduced Spoken Impact Project (SIP) to examine the effect of audio and visual feedback on vocalizations in ASD children. Their experimental results suggested that individual customization is in demand given the children’s varied preferences on different styles of feedback. This inspires us to provide personalized feedback. As for exaggerated feedback, Antti et al.~\cite{granqvist2018exaggeration} contributed a controlled experiment of exaggerating the teaching avatar’s flexibility in a kicking task. The experimental results demonstrate that users prefer exaggerated results over original ones. Our work shares similar insights with their design.

Based on these studies and the previous learning theories on exaggerated feedback as discussed above, we focus on constructing a participatory exaggerated computer-aided pronunciation training system that emphasizes enlarging the teaching effectiveness from a user perception perspective. Specifically, we study how to identify suitable exaggerated feedback to learners with different demands or behaviors in language learning (i.e. different pronunciation proficiencies) and how to define the best set of feedback. Our idea of involving exaggerated feedback and our system of determining the best set of exaggeration parameters can be beneficial for the general area of computer-aided language learning~\cite{garcia2018self, neri2008effectiveness, robertson2018designing, hailpern2009creating} and exaggerated feedback~\cite{granqvist2018exaggeration} systems in HCI.

\subsection{Computer-Aided Pronunciation Training Systems}
\label{sec:2.3}
\noindent\textbf{Development of CAPT.}
Computer-aided pronunciation training (CAPT) system was introduced in the 1960s. The first CAPT system was developed by Kalikow and Swets~\cite{kalikow1972experiments}. They developed a system that used visual feedback to teach English pronunciation for Spanish learners. From 2000 to 2010, most English CAPT systems utilized speech recognition technology, but few of them offered instruction or feedback to learners~\cite{kawahara2004practical}.
From 2010 to the present, researches incorporated diverse technologies into the CAPT system, including pronunciation training method ~\cite{neri2002pedagogy}, Automatic Speech Recognition (ASR)~\cite{neri2006asr}, Mispronunciation Detection and Diagnosis (MDD)~\cite{leung2019cnn}, speech synthesis ~\cite{ning2015hmm, meng2010synthesizing, meng2010synthesizing}, visual-speech synthesis ~\cite{yuen2011enunciate} and application system design~\cite{yuen2011enunciate}. %
Several kinds of online pronunciation corrective feedback in the form of different modalities are proposed in CAPT~\cite{badin2010visual, wong2010development, yuen2011enunciate,liu2012menunciate,brown2020device}, such as audio feedback, text feedback, articulatory feedback~\cite{wong2010development}, etc. Yuen et al.~\cite{yuen2011enunciate} proposed a comprehensive method to produce audio-visual feedback in CAPT. In the next step, they extended their work to a distributed text-to-audio-visual-speech synthesizer (TTAVS) to design a CAPT system with the interactivity on a mobile platform~\cite{leung2013development}. 
Pennington~\cite{pennington1999computer} found that phonology knowledge was not well considered in most CAPT systems. Inspired by the phonology research, our work finds a suitable adjusting range for expressive speech in audio-visual corrective feedback.

\begin{figure*}[t]
  \centering
  \includegraphics[width=0.90\linewidth, height=0.15\linewidth]{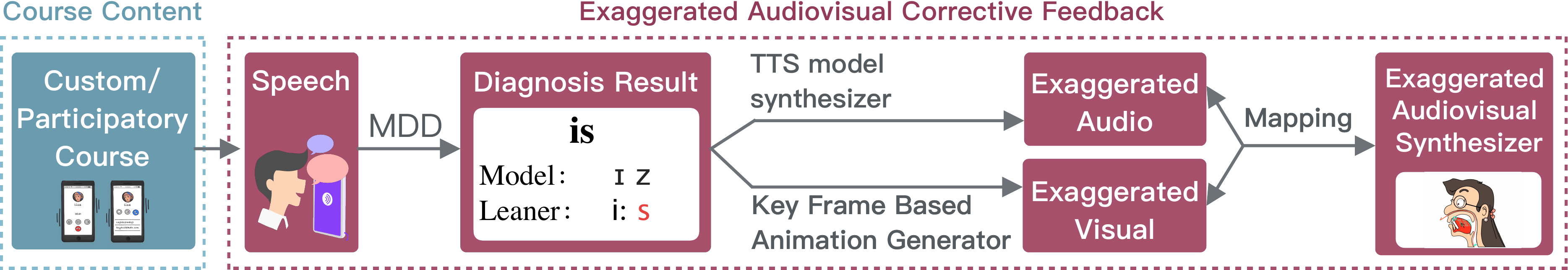}
  \caption{The whole working flow of PTeacher system. It consists of two key components: \textbf{(1)} exaggerated audio-visual corrective feedback generator, and \textbf{(2)} pronunciation training courses. Notably, the system is first served as the platform for user interactions, then the users' feedback also identifies the detailed design of the system.}
  \label{fig:framwork_english}
  \Description{This figure contains 2 part, the first part is the design of the course content, the second part is the mechanism of the exaggerated audiovisual corrective feedback which including MDD and the exaggerated audiovisual synthesizer.}
\end{figure*}

\noindent\textbf{Exaggerated Audio-Visual Feedback in CAPT.}
The above discussed methods fail to increase awareness of learners towards their mispronunciation. Thus, identifiable and perceptible feedback is still urgently needed. 
According to~\cite{ricard1986beyond}~\cite{wohlert2000lip}, in the offline English class, exaggeration is a critical feedback method for the teachers to rectify the pronunciation of learners. Typical exaggerating methods include speaking louder and slower, and showing the movements of mouth clearly to learners. Exaggerations in the form-focused instruction~\cite{wohlert2000lip} have been verified to be beneficial for inexperienced L2 learners. Alghamdi et al.~\cite{alghamdi2017impact} investigated that exaggeration of the visual speech improved the audio-visual recognition of many phoneme classes. Exaggeration methods were used in CAPT systems to assist L2 learners in perceiving stress patterns. Their work provided a fundamental theory, which discussed the effectiveness of exaggeration methods. %
For the exaggerated audio-visual feedback in CAPT,~Zhao et. al~\cite{zhao2013audiovisual} proposed an audio-visual exaggeration method to provide more perceptible corrective feedback. Both exaggerated audio and exaggerated articulatory animation were provided for learners to rectify their pronunciation. %
However, several problems of audio-visual exaggeration in CAPT remain unsolved. 
In our \textit{PTeacher} system, we systematically discuss these problems and present a practicable solution through extensive user studies.

\begin{figure*}[htpb]
  \centering
  \includegraphics[width=0.90\linewidth, height=0.30\linewidth]{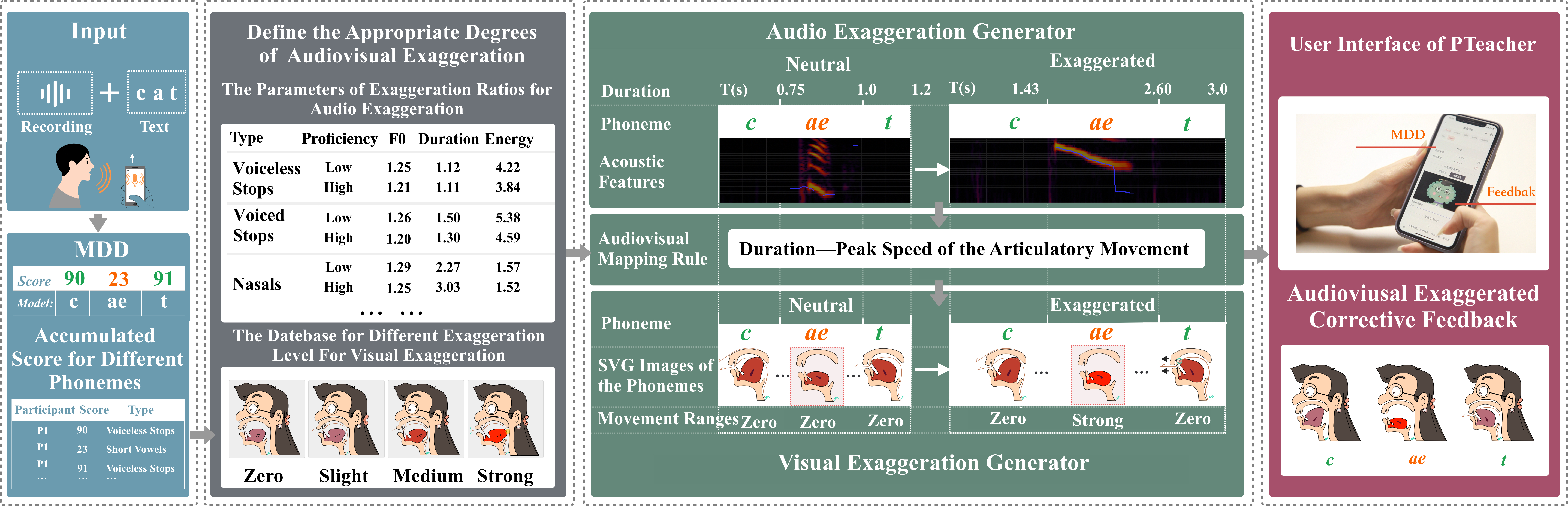}
  \caption{The working flow of the exaggerated audio-visual corrective feedback.
  Mispronunciation detection and diagnosis (\textit{MDD}) systems are employed to diagnose pronunciation mistakes at sentence, word, and phoneme levels.
  Afterwards, personalized feedback are provided by the \textit{audio} and \textit{visual} exaggeration generators based on the MDD results.}
  \label{fig:flower_chart}
  \Description{This figure contains 4 part. The first part is describing the process that the user's speech will deliver to MDD. The second part is audio and visual database. The third part is audiovisual generator. The fourth part is the user interface.}
\end{figure*}

\section{PTeacher System for Pronunciation Training}

In this section, we introduce \textit{PTeacher}, a CAPT system that helps L2 learners correct their pronunciation by incorporating exaggerated audio-visual feedback into a pronunciation course.
The whole pipeline of the system is illustrated in Figure~\ref{fig:framwork_english}. 
It consists of two key components: \textbf{(1)} exaggerated audio-visual corrective feedback generator, and \textbf{(2)} personalized pronunciation training courses. 
Notably, the system is first served as the platform for user interactions, then the users' feedback also identifies the design of the system.

\subsection{Exaggerated Audio-Visual Corrective Feedback}

Exaggerated audio-visual feedback is generated through an audio and a video exaggeration generator based on the different pronunciation situations of each users, which is the key feature of our system.
As shown in Figure~\ref{fig:flower_chart},
Mispronunciation Detection and Diagnosis (\textit{MDD}) systems \cite{li2016mispronunciation,li2015integrating,li2015rating} are firstly employed to detect and diagnose pronunciation mistakes at sentence, word, and phoneme levels.
Afterwards, the \textit{MDD} results are taken by both the \textit{audio} and \textit{visual} exaggeration generator to generate exaggerated feedback.

\subsubsection{Accumulated Pronunciation Diagnosis}
\label{sec:3.1.1}
Different from previous systems \cite{zhao2013audiovisual}, the historical \textit{MDD} results are also considered to determine the exaggeration level. The \textit{MDD} results are exponentially decayed and accumulated using the following equation:
\begin{align}
R=(1-\alpha)^{n}R_{n}+\sum_{0\leq i\leq n-1}\alpha(1-\alpha)^{i}R_{i}
\end{align}
where $n$ is the number of all results, $R_0$ is the latest result, $R_i$ is the $i$-th historical result and $\alpha$ is the decay ratio. In our research, $\alpha$ is set to $0.9$.
After accumulated, the phoneme with the lowest score in one sentence will be selected and exaggerated by the audio and video exaggeration generators.

\subsubsection{Audio Exaggeration Generator.}
\label{sec:speech}
\begin{figure*}[htpb]
  \centering
  \includegraphics[width=0.9\linewidth, height=0.185\linewidth]{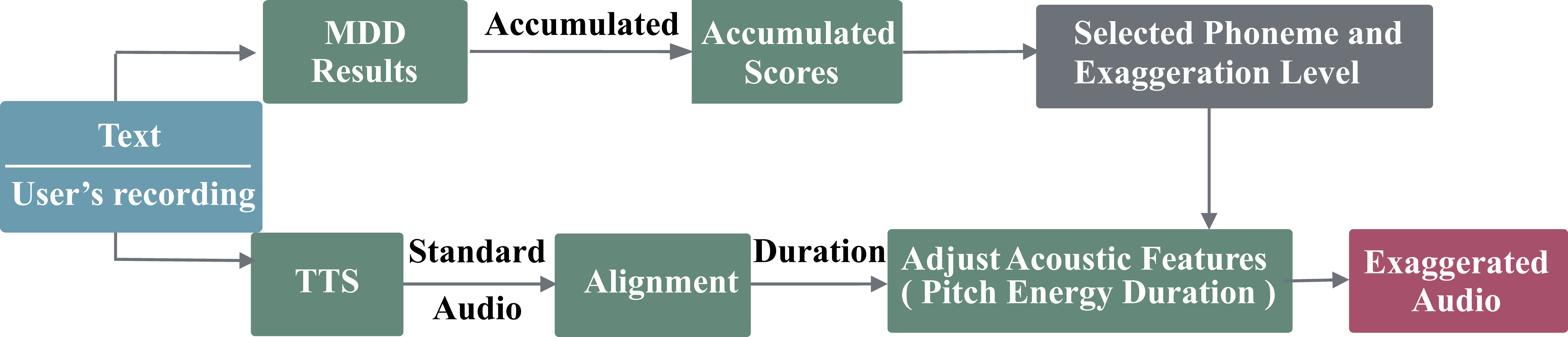}
  \caption{The working flow of the audio exaggeration generator. Text-To-Speech (TTS) model is used to synthesize neutral speech with given text.
  Montreal Forced Alignment~(MFA) algorithm is leveraged to locate the position of the selected phoneme in the synthesized speech.
  The exaggeration level is determined by the accumulated score in the \textit{MDD} results.
  Then pitch, duration and energy of the selected phoneme are exaggerated with the parameters of the corresponding exaggeration level with PyWorld.
  }
  \Description{This figure describes the methods of how to generate exaggerated audio, which contain 3 part including user input, TTS and the exaggerated audio.}
  \label{fig:audio_exaggeration}
\end{figure*}

The framework of the audio exaggeration generator is depicted in Figure~\ref{fig:audio_exaggeration}.
The exaggerations are generated by applying adjustment of proper exaggeration level to synthesized speech based on the \textit{MDD} result.

A pre-trained Text-To-Speech~(TTS)~model \cite{ren2019fastspeech} is used to synthesize high-quality, neutral speech with the given text.
Montreal Forced Alignment~(MFA)~\cite{mcauliffe2017montreal} algorithm is leveraged to locate the position of the selected phoneme in the synthesized speech.
Then pitch, duration and energy of the selected phoneme are exaggerated with the parameters of the corresponding exaggeration ratios with PyWorld \cite{hsupython}.
The exaggeration level for the selected phoneme is determined by the accumulated score as described in Section~\ref{sec:3.1.1}.
Two exaggeration ratios are used in our research.
The scores in $[0,50)$ are projected to the "Low Proficiency" exaggeration ratios
and scores in $[50,100]$ are projected to the "High Proficiency" exaggeration ratios.
The parameters for each exaggeration ratios are determined by personalized audio exaggeration experiment~\ref{sec:exp_audio}.

\begin{figure*}[htpb]
  \centering
  \includegraphics[width=0.9\linewidth, height=0.45\linewidth]{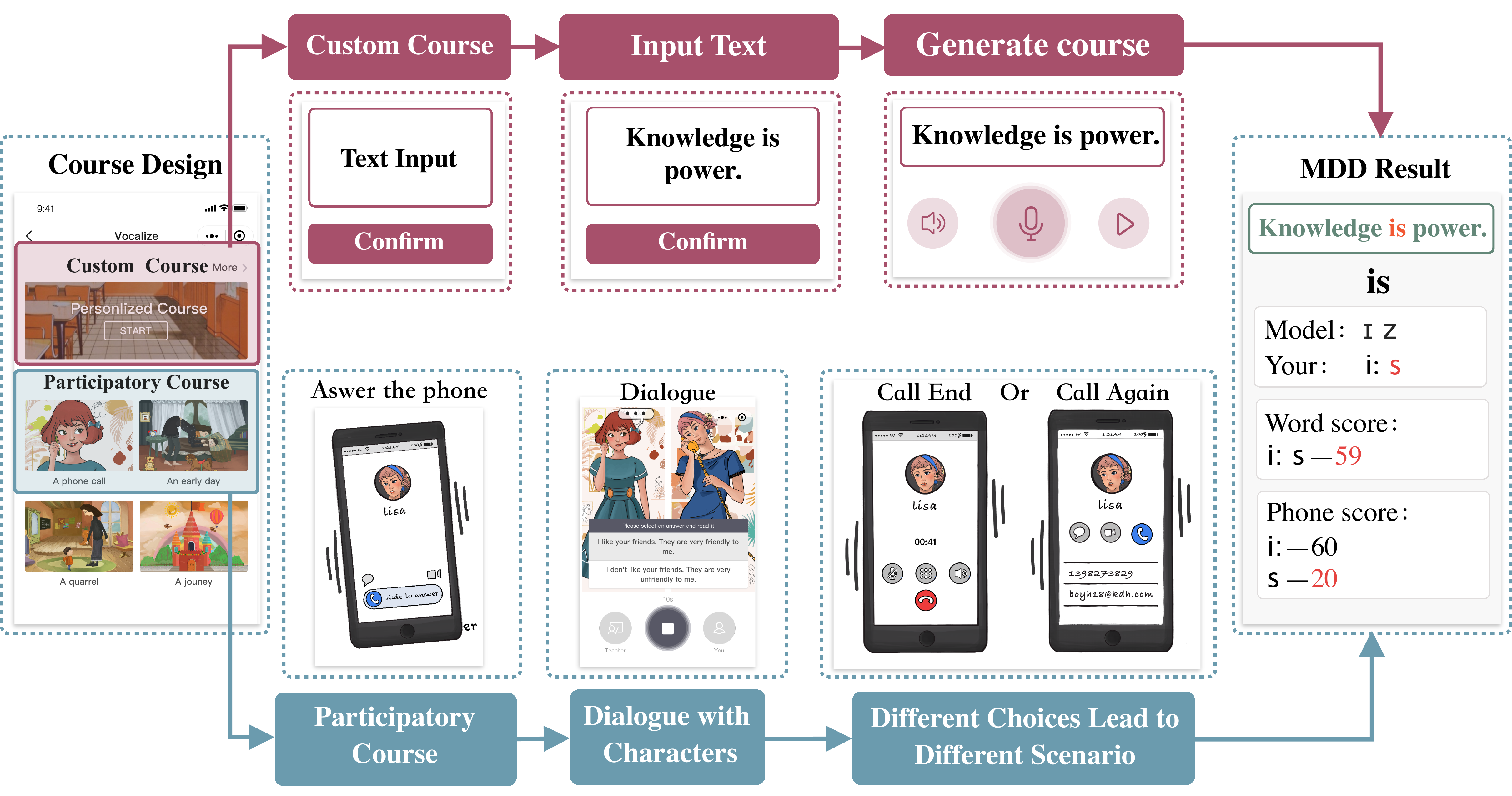}
  \caption{The workflow of the interactive courses and the custom course. In the interactive courses, user learn the drama courses. In the custom course, the system will generate a flexible course according to the English text which the user input.}
  \label{fig:course}
  \Description{This picture illustrates the interface of course design. In the interactive courses, there are a number of interactive courses for learner to communicate. In the custom course, the system will automatically generate a course according to the learner's preferences. }
\end{figure*}

\begin{figure}[ht]
  \centering
  \includegraphics[width=1.0\linewidth]{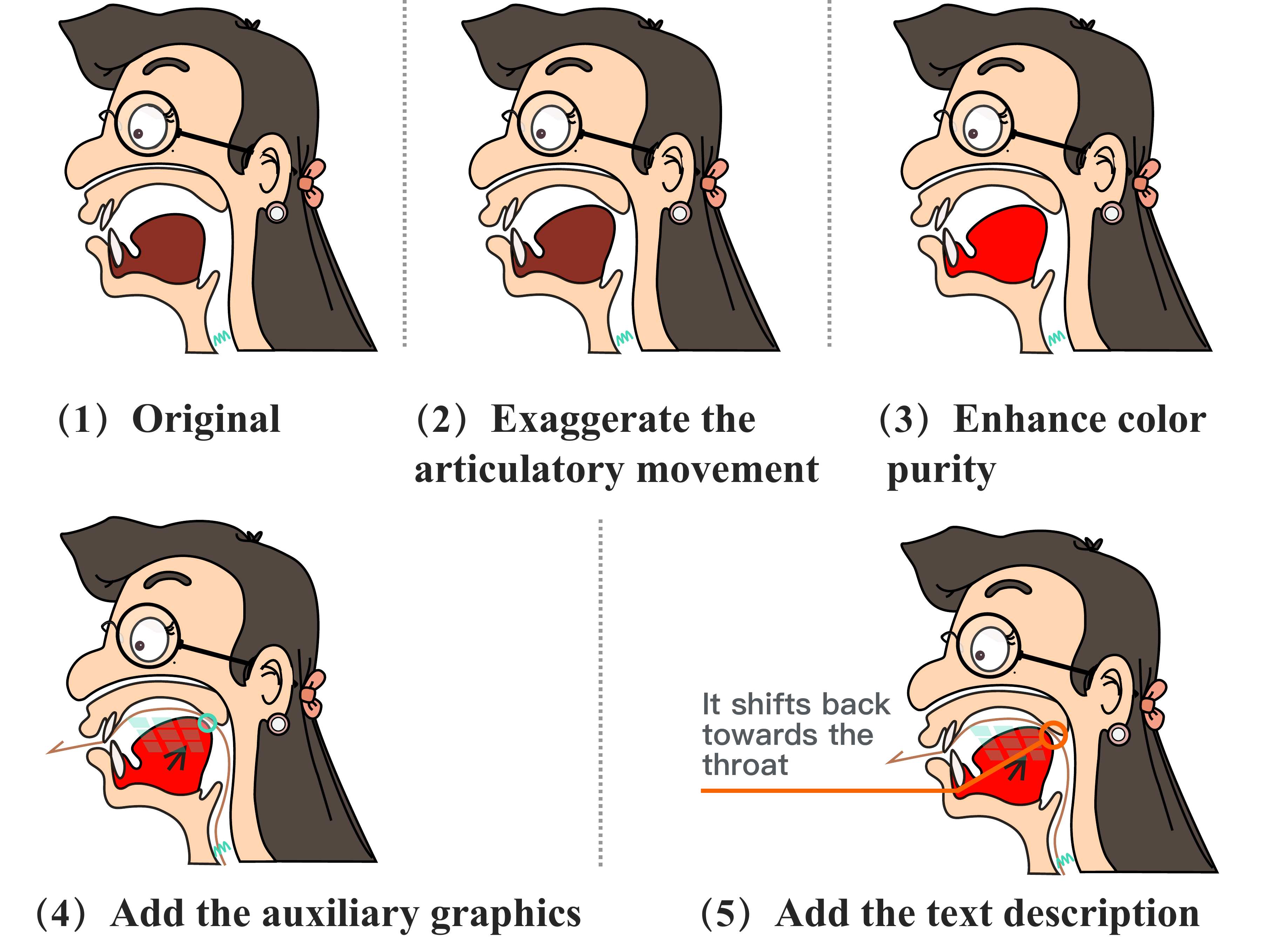}
  \caption{We exaggerate the mouth movement, the color of the key organs and add auxiliary graphics as well as multiple text descriptions.}
  \label{fig:visual_example}
  \Description{This picture shows the examples of how to exaggerate an articulatory animation.}
\end{figure}

\subsubsection{Visual Exaggeration Generator.}
\label{sec:visual}
We adopt the design of articulatory animation ~\cite{lasseter1987principles,chang1993animation,ribeiro2012illusion} to provide visual exaggeration and further increase the ability of expression in the visual domain.
Our visualization plots three components: articulatory movement, tongue color, and auxiliary sign. %
For the articulatory movement, we plot the key parts of the articulatory actions~(oral cavity), while irrelevant parts (\textit{oesophagus, epiglottis, nasal, etc.}) are simplified. 
Four degrees of exaggerated side-view and front-view viseme components are designed under the guidance of articulatory phonemeticians and animation designer.
Having obtained the articulatory plots of each phoneme, we leverage the Viseme Blending~\cite{ezzat2000visual} to interpolate the overall animation. 

We generate the exaggerated articulatory animation using the following methods: (1) increase the amplitude of key articulatory movement. (2) We modify the color of the tongue from low purity to high purity when exaggeration is needed, to draw the attention of learners.
(3) Auxiliary graphics (\textit{e.g. arrows, airflow}) and supplemental texts (\textit{e.g. manner of articulation}), are finally added to help learners better understand the pronunciation through visualization, as shown in Figure~\ref{fig:visual_example}. 
It is noteworthy that we update the design in the amplitude of articulatory movements according to human interactions. Different degrees of exaggerated will also be provided to different users in a personalized manner. Details are illustrated in Section \ref{sec:exp6}. %

\subsection{Personlized Course Content Design}
\label{sec:course}
We propose to personalize the pronunciation training through our novel course design to build the connection between general CAPT systems and individual users. 
As shown in figure~\ref{fig:course}, two types of courses, namely \textit{Custom Course} and \textit{Participatory Course} are designed.

The \textit{Participatory Course} is actually a particular type of Interactive Participatory Drama (IPD)~\cite{hubbard2002interactive} in language learning. In this type of course, learners play active roles~\cite{bu2018icoobook} in pre-programmed scenarios by haptic and voice interaction (such as chatting, painting, etc). The course includes multiple storylines, depending on the choice of learners in the story.
The primary purpose is to stimulate learners to do more perception~(hear) and productive training~(speak) in a participatory manner.
While in \textit{Custom Course}, learners could customize the content of the lessons based on their particular needs.

Both courses take recorded audios from users as inputs, and leverage the aforementioned exaggerated audio-visual feedback for users to recognize and rectify their mistakes. The learners' proficiencies are determined by \textit{MDD} scores. The system assesses users’ pronunciation accuracies on 14 types of phonemetic symbols and makes a comprehensive judgment according to the user's immediately performance and historical proficiency. Thus personalized feedback can be dynamically generated in a life-long manner. 
We expect users to actively participate in pronunciation learning by noticing their improvements through reading the comprehensive reports from \textit{PTeacher}.

\section{User-Participated Experiments}
In general, with the users-participated experiments, we determine the optimum audio exaggeration ratios, find the most appropriate visual exaggeration level and verify the effectiveness of \textit{PTeacher}. Participant are interviewed during the experiments and grounded theory approach~\cite{corbin2014basics} is leveraged to analyze the interview data. 
\begin{itemize}
    \item In audio experiments, we firstly define four audio exaggeration levels and determine the exaggerations ratios for four levels (Section~\ref{sec:audio_material_preparation}). 
    For each audio exaggeration level, we evaluate a) Distinguishability (Section~\ref{sec:distinguishability}); b) Understandability (Section~\ref{sec:understandability}); and c) Perceptibility (Section~\ref{sec:perceptibilty}). 
    Based on the result of the experiments mentioned above, we apply nonlinear fitting to determine the optimum exaggeration ratios (Section~\ref{sec:optimum_ratios}).
    \item In visual experiments, we first define multiple exaggeration levels for articulatory movements and tongue colors (Section~\ref{sec:visual_preparation}). 
    Then, we find the optimum exaggeration level with the highest user perceptibilty rate (Section~\ref{sec:visual_test}). 
    \item In supplementary experiments, we evaluate the effects of user engagement and user experience between participatory course and custom course which is described in the supplementary materials (Chapter 1).
    \item Finally we verify the effectiveness of our system by comparing training effects of our system with other systems (Section~\ref{sec:verify}).  
\end{itemize}

\subsection[Participants]{Participants}
30 L2 learners,including 15 high proficiency learners and 15 low proficiency learners are invited to participate two audio exaggeration experiments which are audio distinguishability experiment (Section~\ref{sec:distinguishability}) and understandability experiment (Section~\ref{sec:understandability}). %
The ages of 30 learners range from 20 to 32. %
Then 22 native English speakers together with 30 L2 learners in the previous experiments are invited to participate in the audio perceptibility experiment (Section~\ref{sec:perceptibilty}). %
Among 22 native speakers,
10 are from South Africa,
5 are from the United States,
4 are from the United Kingdom
and 3 are from Canada. %
The ages of the native speakers range from 24 to 42. %
All of the native speakers are certificated with TESOL (Teaching English to Speakers of Other Languages) issued by Ascentis, which is an officially recognized TESOL certificate authority. %
The 30 L2 learners from the previous experiments are further invited to participate in the visual perceptibility experiment (Section~\ref{sec:visual_test}). %
Then, we invite 20 new L2 learners to take part in the supplementary experiments.
80 L2 learners, including 50 new L2 learners, and 30 learners in the previous experiment, are invited to participate in the \textit{PTeacher} effectiveness verifying experiment (Section~\ref{sec:verify}). %
The 80 learners contain 40 high proficiency learners and 40 low proficiency learners. %
The ages of 80 learners range from 20 to 32. %

In each experiment, 20\% of the participants are randomly selected to participate in face-to-face interviews based on the following criteria: (a) the ratio of low proficiency learners to high proficiency is 1:1; (b) the ratio of male to female is 1:1; (c) the participants are aged between 20 and 32. We choose 20 participants with an average age of 24. %
Grounded theory approach~\cite{corbin2014basics} is conducted with interview data analysis.
All interview data are recorded by Google doc and processed by MaxQDA3~\cite{oliveira2013thematic} for qualitative analysis.
Through open coding, 100 codes~\cite{liu2019bought} are produced.
We collaboratively synthesized the interview content into higher-level themes through axial coding~\cite{liu2019bought}, including learning challenges, learning experiences, learning efficiency and learning effects.
We also discussed the internal connections between these themes and generated an interviewer report.

\subsection[Experiments on Audio Exaggeration]{{Experiments on Audio Exaggeration}}
Audio exaggeration experiments are conducted to optimize the exaggeration ratios of the audio exaggeration generator in terms of distinguishability, understandability and perceptibility.
We first synthesize the experimental audios with four exaggeration levels based on exaggerated speeches from a pronunciation training expert.
Then we test distinguishability, understandability and perceptibility for learners in the low proficiency and high proficiency groups w.r.t. each exaggeration level.
Based on the results, we apply a non-linear fitting to determine the optimum exaggeration ratios for learners with low proficiency and high proficiency.

\subsubsection{Material Preparation}
\label{sec:audio_material_preparation}
\label{sec:exp_audio}
First, a pronunciation training expert is asked to read 350 representative words from the Oxford Dictionary with four exaggeration levels: \textit{zero}, \textit{slight}, \textit{medium} and \textit{strong}, respectively.
We calculate the exaggeration ratios of the pitch, duration and energy for each exaggerated phonemes.
The expert then adjusts the exaggeration ratios. 
The exaggeration ratios for each exaggeration level w.r.t. different type of phonemes are shown in the supplementary material.
Finally, we synthesize 800 speeches with different exaggeration level as test materials.

\begin{figure*}[htpb]
  \centering
  \includegraphics[width=0.90\linewidth, height=0.29\linewidth]{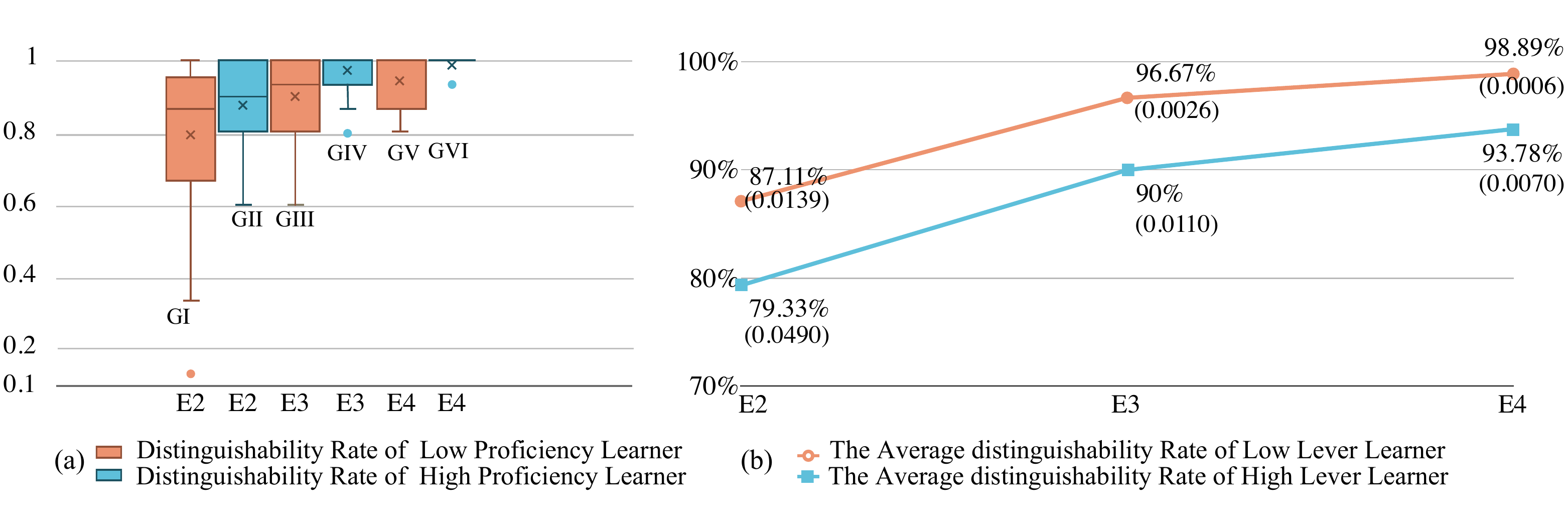}
  \caption{The distribution and average of the distinguishability rates are shown in figure (a) and (b), respectively. The variances of the distinguishability rates is in the brackets in figure (b). E1, E2, E3 and E4 represent \textit{zero}, \textit{slight}, \textit{medium} and \textit{strong} audio exaggeration respectively.}
  \label{tab:Distinguishablility_figure}
  \Description{The figure shows the distinguishability rates of different exaggeration levels for learners with different degree of proficiency.}
\end{figure*}

\subsubsection{Experiment on Distinguishability of Exaggerated Speech}
\label{sec:distinguishability}
Distinguishability rate~\cite{liberman1957discrimination}, which indicates whether a listener can easily distinguish the exaggerated phoneme in the exaggerated audio, is a crucial evaluation index to verify the effect of exaggerated expression.
In our case, given a speech with one exaggerated phoneme, it evaluates whether the participants can discern the exaggerated phoneme from the speech.
The distinguishability rate is defined as the accuracy of whether the user can discern the exaggerated phoneme.

The distribution and average of distinguishability rates are shown in Figure~\ref{tab:Distinguishablility_figure}. %
It demonstrates that higher exaggeration level produces higher distinguishability rate. %
The average distinguishability rate increases from 87.11\% to 98.89\% for learners with high proficiency. %
The average distinguishability rate increases from 79.33\% to 93.78\% for learners with low proficiency. %
The result also indicates that learners with lower proficiency need audio with higher exaggeration level to discern the exaggerated part as easily as those with higher proficiency. %
We carry out single tail t-test between groups of the same proficiency with different exaggeration levels and different proficiency with different exaggeration levels. %
The P results mainly range from 0.0013 to 0.013, indicating significant differences. T-test result P between test groups with \textit{medium} and \textit{strong} exaggeration for low proficiency learners reaches 0.0678. The value is acceptable since it is close to 0.05. %

The user interview also demonstrates the experiment result. %
A learner with low proficiency says \emph{``Audio exaggeration helps me locate where I need to pay attention. It it was quite easy for me to locate the exaggerated part for level 3 and level 4 exaggerated audios."} %

\begin{figure*}[h]
  \centering
  \includegraphics[width=0.86\linewidth, height=0.28\linewidth]{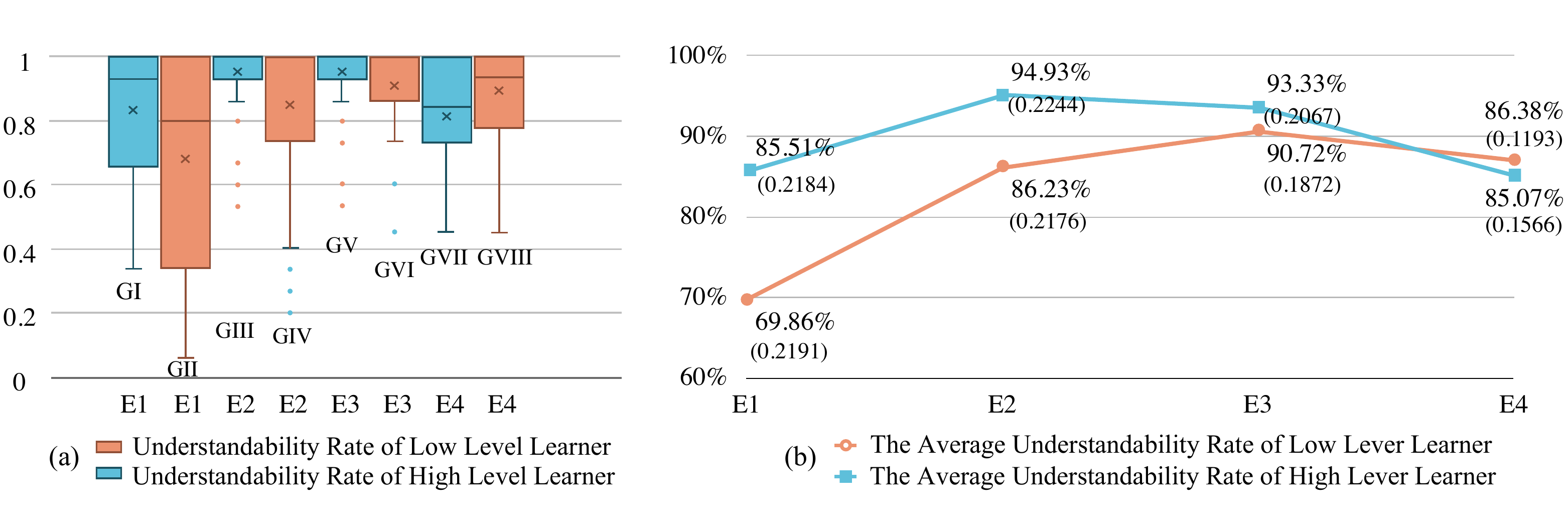}
  \caption{The distribution and average understandability rates are illustrated in figure (a) and (b), respectively. The variance of the understandability rates is in the brackets in figure (b).}
  \label{tab:udsp}
   \Description{The figure shows the understandability rates of different exaggeration levels for learners with different degree of proficiency.}
\end{figure*}

\subsubsection{Experiment on Understandability of Exaggerated Speech}
\label{sec:understandability}

Understandability rate indicates whether the learner can still easily understand the exaggerated speech without confusion~\cite{veselovska2016teaching, munro1999canadians, cervino2010investigating, mora2017relationship}.
Given two similar phoneme and a speech with one of the phonemes being exaggerated, participants are asked to choose which phoneme appears in the word. %
Similar to distinguishability rate, understandability rate is defined as the accuracy of whether the user can still recognize the exaggerated phoneme. %

The distribution and average of the understandability rates are illustrated in Figure~\ref{tab:udsp}. Unlike the result of the experiment on distinguishability, a higher exaggeration level does not lead to the increase of understandability rates. %
For the learners with high proficiency, the optimum exaggeration level is \textit{slight}, with the highest understandability rate at 94.93\%. %
For learners with low proficiency, the optimum exaggeration level is \textit{medium}. The highest understandability rate is 90.72\%. %
With the \textit{strong} exaggeration, the average understandability rates drop to 85.07\% and 86.38\% for high and low proficiency learners, respectively. %
The reason is that \textit{strong} exaggeration can cause distortion, which impedes learners from understanding it correctly, especially for learners with higher proficiency. %
We carry out single tail t-test between groups of the same proficiency with different exaggeration levels and different proficiency with different exaggeration levels. %
Most of the P-values range from 0.0017 to 0.0086, indicating a significant difference. %
The resulted P between the group with \textit{medium} and \textit{slight} exaggeration for high proficiency learner is 0.2480, indicating almost no difference. %
Also, P-value between the group with \textit{medium} and \textit{slight} exaggeration for low proficiency learner is 0.1198, indicating a minor difference. %
Based on the t-test results, we confirm that both \textit{medium} and \textit{slight} exaggeration is acceptable for learners with different proficiencies. %

The participant interview further also demonstrates part of the experiment results. %
A learner with high proficiency comments that \emph{``The voices break in some E4 exaggerated stops, and the phoneme `b' sounds like `p'. "}
Another learner with high proficiency comments that \emph{``The `r' sounds very strange in E4. I cannot tell you what it is."} %
A learner with low proficiency comments that \emph{``I cannot recognize the `th' sound in audio with level two exaggeration, because `th' and `sh' are just too similar."} %

\begin{figure*}[htpb]
  \centering
  \includegraphics[width=0.90\linewidth, height=0.28\linewidth]{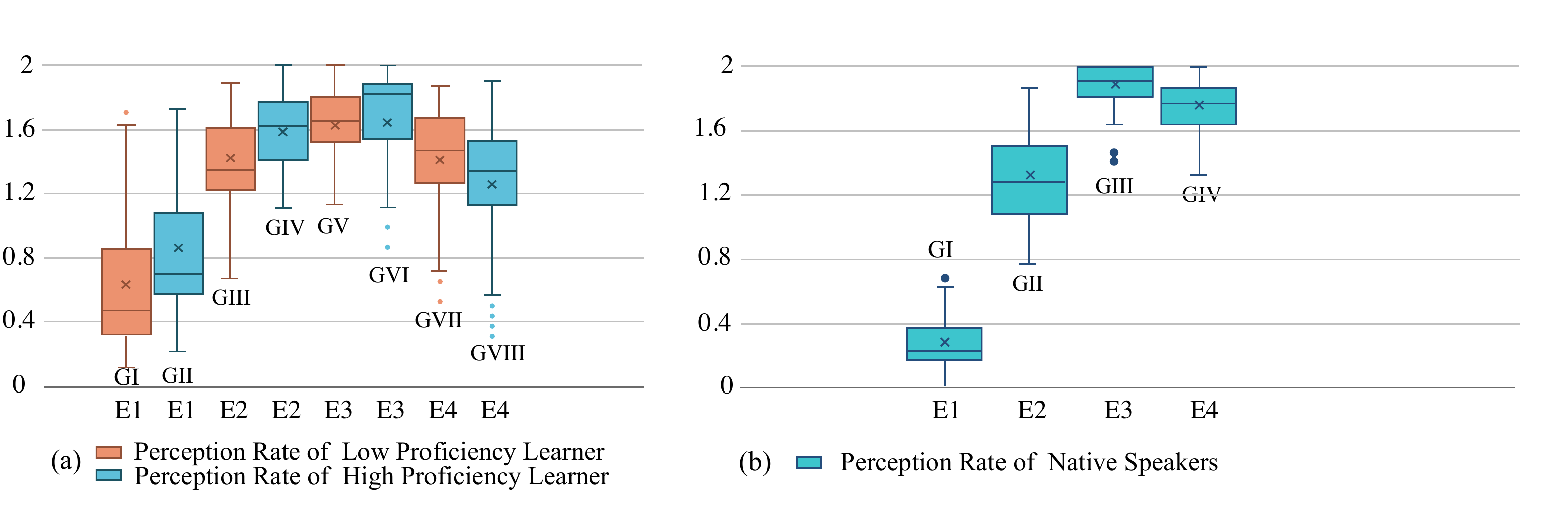}
  \caption{The perceptibility rates distribution w.r.t. four exaggerations for the learner with different proficiency are depicted in figure (a). The distribution of perceptibility rates assessed by the teacher is shown in the figure (b).}
  \label{tab:pds}
  \Description{The figure shows the auditory perceptibility rates of different exaggeration levels for learners with different degree of proficiency.}
\end{figure*}

\subsubsection{Experiment on Perceptibility of Exaggerated Speech}
\label{sec:perceptibilty}
Perceptibility score, which evaluates exaggerated speech's perceptual quality, is a significant indicator to check the intensity level~\cite{zeng2000auditory} of perception from the perspective of hearing. 
Participants are asked to give opinion scores ranging from 0 to 5 (0 for too weak to perceive, 5 for too strong) in terms of perceptual exaggeration level. %
The perception scores are calculated with:
\begin{align}
P=2.5-|2.5-Score|,
\end{align} 
where Score is the opinion score and P is the perception scores. %

The distribution of perceptibility scores for different exaggeration level is illustrated in Figure~\ref{tab:pds}. %
We find that the distribution is similar to that of the understandability rates. This result indicates that neither \textit{slight} exaggeration nor \textit{strong} exaggeration is good enough.

The participants' interview further confirms our inference. A learner with low proficiency comments that:\emph{``I felt that the 'slight' exaggerations are so slight sometimes that I have difficulties realizing them."} %
A learner with high proficiency comments that: \emph{``I don't think the slight exaggeration is good enough since it is too weak. Also, I found severe distortions in the 'strong exaggeration' version. So I don't think it is good enough, either."} %
A native speaker comments:\emph{``I think the medium exaggeration version is very cool. I must have used a similar exaggeration method in my class."} %

\subsubsection{Optimizing Exaggeration Ratios with Non-linear Fitting}
\label{sec:optimum_ratios}
The experiments on distinguishability rates, understandability rates, and perceptibility scores apply a non-linear fitting to find the optimum exaggeration ratios w.r.t. different phoneme types. The optimum exaggeration ratios can achieve the highest value defined with:
\begin{align}
V = DR + UR +\frac{PS}{2.5},
\end{align}
where $V$ is the optimizing target for non-linear fitting, $DR$, $UR$ and $PS$ are distinguishability rates, understandability rates and perceptibility scores respectively. %

\begin{figure*}[htpb]
  \centering
  \includegraphics[width=0.9\linewidth, height=0.32\linewidth]{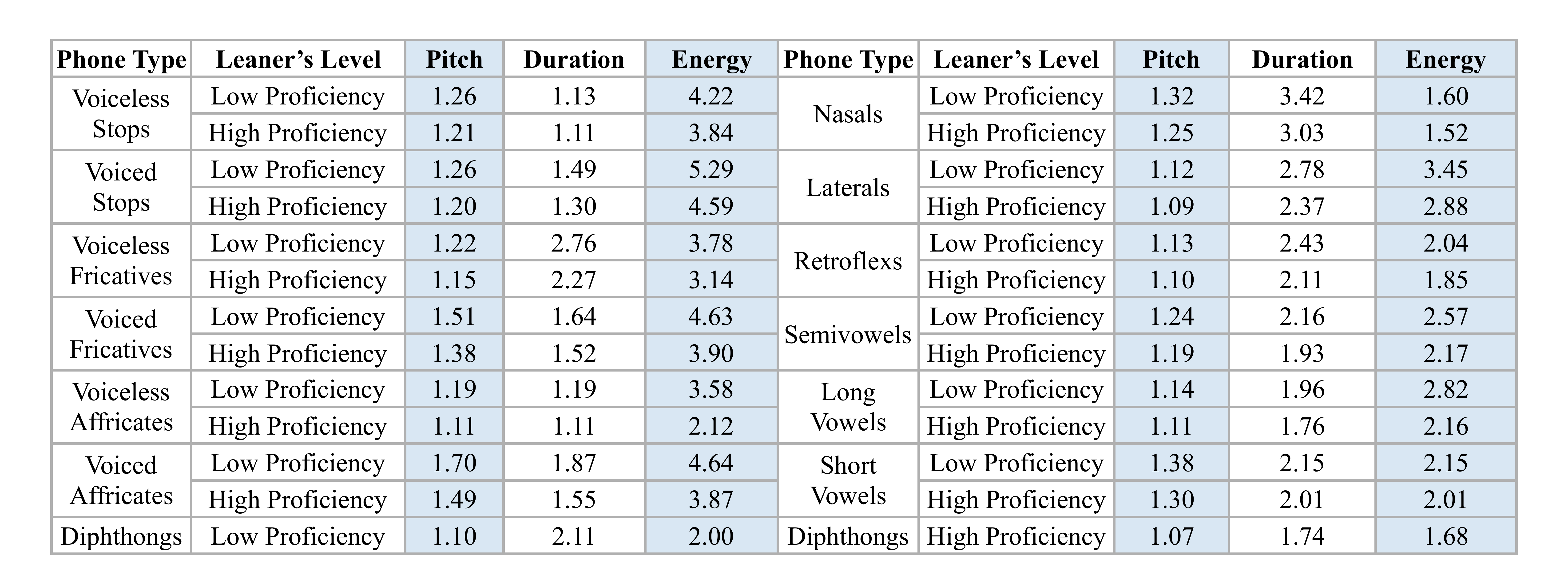}
  \caption{The optimum audio exaggeration ratios are shown in the figure. We apply non-linear fitting to find the exaggeration ratios with the highest distinguishability rates, understandability rates and perceptibility rates.}
   \Description{The figure shows the optimum audio exaggeration ratios with the highest distinguishability rates, understandability rates and perceptibility rates. For example, The exaggeration ratios of the Voiceless stops' pitch is 1.26 for the low proficiency learners.}
  \label{tab:best audio rate}
\end{figure*}

The result is shown in Figure~\ref{tab:best audio rate}. %
Almost all the best exaggeration ratios fall into the range of medium exaggeration. %
We notice that the exaggeration ratios needed for high proficiency learners are lower than those low proficiency learners. %
The optimum exaggeration ratios are used in the following experiments.

\begin{figure}[h]
  \centering
   \includegraphics[width=0.9\linewidth, height=0.65\linewidth]{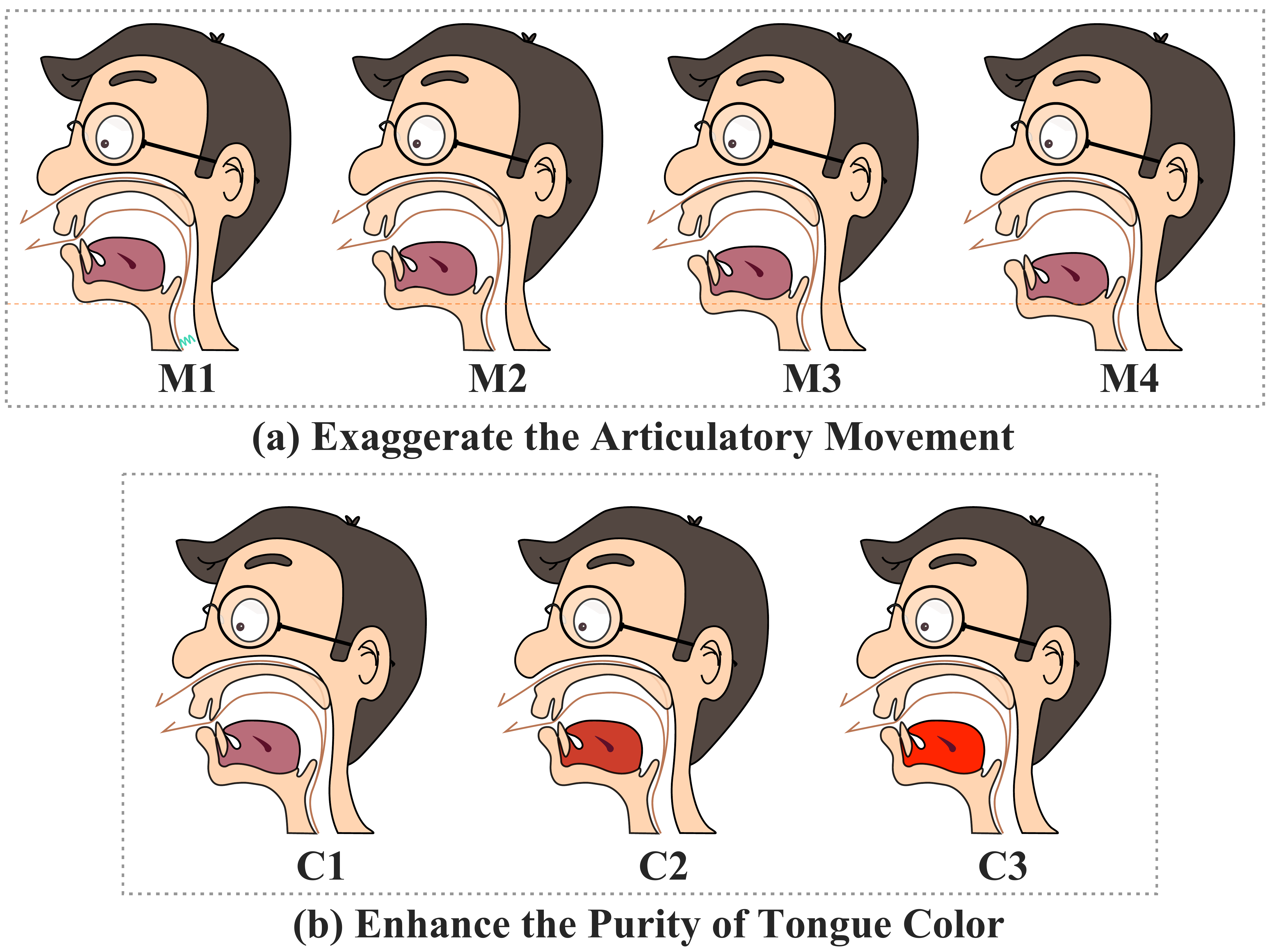}
   \caption{The figure illustrates examples of animations with exaggeration. M1, M2, M3 and M4 represent zero, slight, medium and strong articulatory movement exaggeration, respectively. C1, C2 and C3 represent zero, slight and strong tongue color exaggeration, respectively.}
   \Description{The figure contains two part, the first part shows the amplitude of the movement of articulatory organs become more and more larger, the second part shows the purity color of tongues become more and more higher. }
   \label{fig:visual_example1}
\end{figure}

\subsection[Experiments on Visual Exaggeration]{{Experiments on Visual Exaggeration}}

\label{sec:exp6}

Visual exaggeration experiments are conducted to optimize the articulatory movement exaggeration level and tongue color exaggeration level in terms of perceptibility. 
We first synthesize the experiment video with different articulatory movements exaggeration level and tongue color exaggeration level. 
Then we test perceptibility for learners in the low proficiency and high proficiency groups w.r.t. different exaggeration level.
Based on the results, we determine the optimum articulatory movement exaggeration level and tongue color exaggeration level for the learners with low proficiency and high proficiency.

\subsubsection{Material Preparation for Visual Exaggeration}
\label{sec:visual_preparation}
Four exaggeration levels of articulatory movement, which are named \textit{zero}, \textit{slight}, \textit{medium}, and \textit{strong}, are manually designed. %
Three exaggeration level of tongue colors, which are named \textit{zero}, \textit{medium}, and \textit{strong} respectively, are also manually designed. %
We synthesize the animations of 52 words, equally covering the 13 phoneme types, with different exaggeration levels w.r.t. articulatory movement and tongue color. %
An example of the animations is shown in Figure~\ref{fig:visual_example1}. %

\begin{figure*}[htpb]
  \centering
   \includegraphics[width=0.90\linewidth, height=0.48\linewidth]{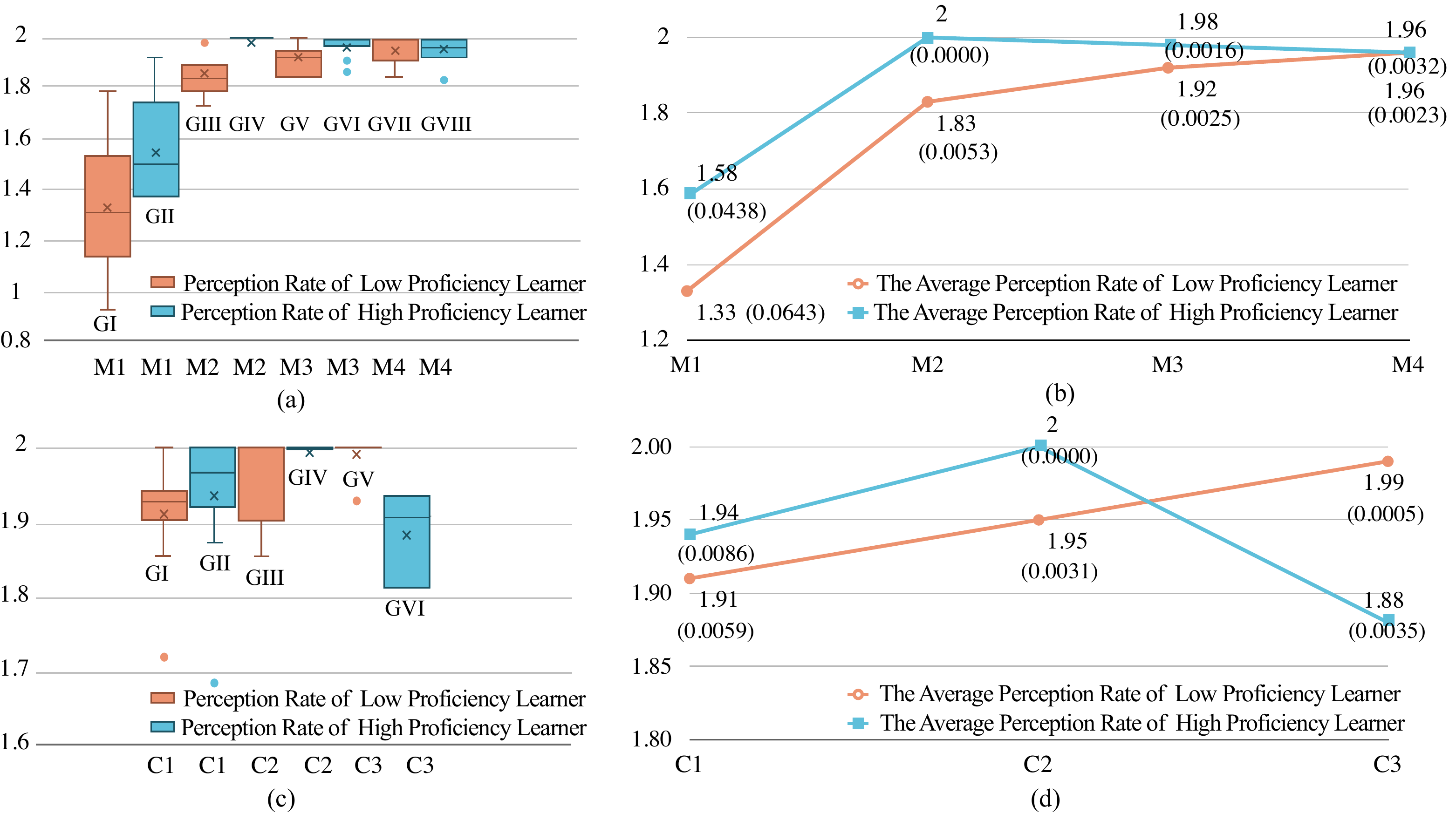}
   \caption{The distribution and average of the perception rates for different articulatory movement exaggerations are shown in figure (a) and figure (b), respectively. The distribution and average of the perception rate for different tongue color exaggerations are shown in figure (c) and figure (d), respectively.}
   \Description{The figure shows the visual perceptibility rates of different exaggeration levels for learners with different degree of proficiency.}
   \label{tab:allf_v}
\end{figure*}

\subsubsection{Experiment on Perceptibility of Exaggerated Animations}
\label{sec:visual_test}
We also conduct perceptibility experiment on animations with visual exaggeration. %
We directly take the definition of opinion score and perceptibility score in the experiment on audio exaggeration (Section~\ref{sec:perceptibilty}).
Participants are first asked to give opinion scores on animations with zero tongue color exaggeration and different articulatory movement exaggeration. 
Then, we find optimum articulatory movement exaggeration with the highest perceptibility score. %
Following that, participants are asked to give opinion scores on animations with optimized articulatory movement exaggeration and different exaggeration. %
Finally, we find optimum tongue color exaggeration with the highest perceptibility score. %

The distribution and average of perceptibility scores on different articulatory movement exaggeration and tongue color exaggeration are shown in Figure~\ref{tab:allf_v}. %
We find that a higher articulatory movement exaggeration level or tongue color exaggeration level does not lead to the increase of understandability rates. %
The optimum articulatory movements exaggeration level is a \textit{slight} exaggeration for learners with high proficiency, \textit{strong} exaggeration for learners with low proficiency. %
The perceptibility scores are 2 and 1.96, respectively. %
The optimum tongue color exaggeration is \textit{slight} for learners with high proficiency. They obtain a perceptibility score of 2. %
While the optimum tongue color exaggeration for learners with low proficiency is a \textit{strong} exaggeration. The perceptibility score is 1.99. %
We carry out single tail t-test between groups of the same proficiency with different exaggeration levels and different proficiency with different exaggeration levels. %
Most of the P-values range from 0.00011 to 0.037, indicating a significant difference. %
The t-test result P between the group with \textit{medium} and \textit{strong} articulatory movement exaggeration for low proficiency learner is 0.1466, indicating almost no difference. %
Also, P-value between the group with slight and strong tongue color exaggeration for high proficiency learner is 0.1378, indicating a slight difference. %
Based on the t-test result, we find that both \textit{strong} and \textit{medium} articulatory exaggeration is acceptable for learners with low proficiency. %
We also see that both \textit{slight} and \textit{strong} articulatory exaggeration is acceptable for learners with high proficiency. %
Still, we choose the optimum articulatory movement exaggeration and tongue color movement mentioned above as the visual exaggeration setting in the following experiments. %

The participant interview further confirms our conclusion. A learner with low proficiency comments: \emph{``I cannot see any different with the low exaggeration level on articulatory movement. Same to the tongue color exaggeration. It is not very eye-catching."} A learner with high proficiency comments:\emph{``The high-level exaggeration may be too much for me."}

\subsection[Experiment on Effectiveness of PTeacher]{ Experiment on Effectiveness of PTeacher}
\label{sec:verify}
\label{sec:exp10}
To determine if the personalized feedback is effective, we compare our system to 3 other systems.
The first system is set with no exaggeration. %
The second system is set with no personalization, which means the exaggeration ratios are fixed to the average of the exaggeration ratios for low proficiency and high proficiency. %
The third system uses feedback from pronunciation training experts. %
Participants are equally divided into four groups and are asked to test their pronunciation accuracy before and after one-hour training with different systems. %
Their pronunciation accuracy is annotated by pronunciation training experts with percentile.
The improvement rate of learners is calculated with:
\begin{align}
I=\frac{S_{\text{after}}-S_{\text{before}}}{S_{\text{before}}}
\end{align}
where $I$ is the improvement rate, $S_{\text{before}}$ and $S_{\text{after}}$ is pronunciation accuracy of the learner before and after training respectively. %

\begin{figure*}[h]
  \centering
  \includegraphics[width=0.90\linewidth, height=0.28\linewidth]{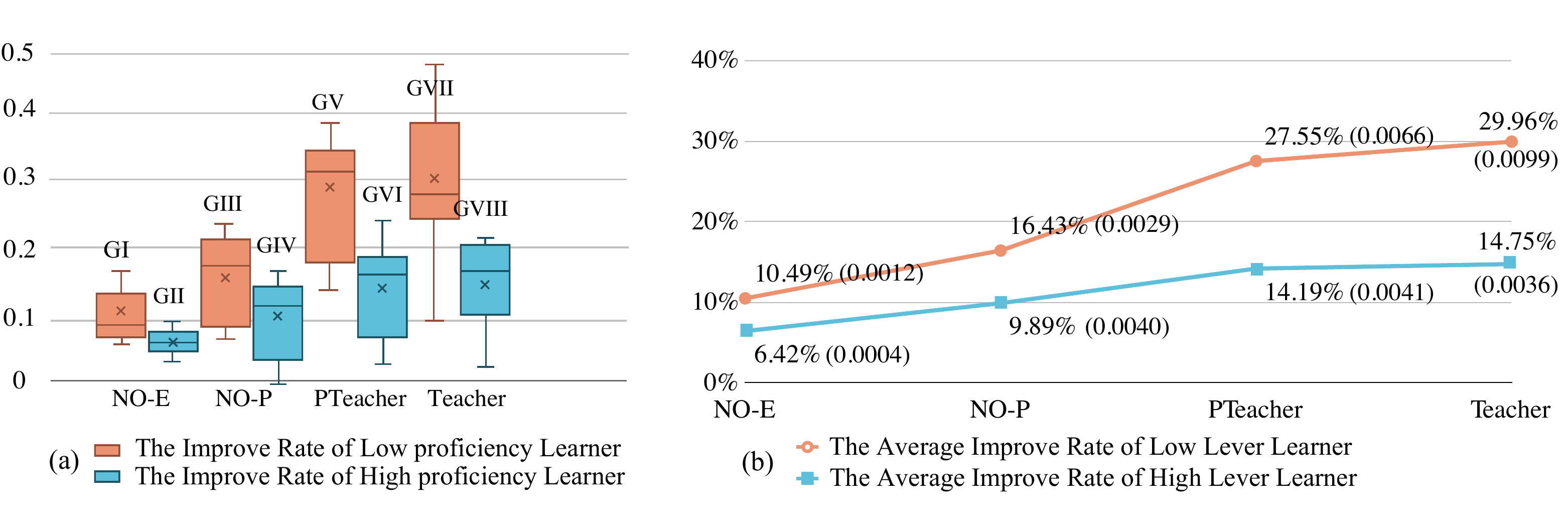}
  \caption{The distribution and the average of the improvement rates are illustrated in figure (a) and (b), respectively. The NO-E, NO-P, \textit{PTeacher} and Teacher represent the system (1) without exaggeration feedback (2) without personalizing (3) personalized exaggeration feedback (4) pronunciation training experts, respectively. \textit{PTeacher} is almost as efficient as teacher-aid training, far exceeding systems without personalized exaggeration feedback.}
 \Description{This figure show the improvement rate of different English proficiency learners after a hour of pronunciation training with PTeacher and the other 3 system}
  \label{tab:finalf}
\end{figure*}

The result is shown in Figure~\ref{tab:finalf}. %
We find that the improvement rate of learners who are trained with \textit{PTeacher} is much higher than those who are trained with the non-exaggeration system or the non-personalizing system is comparable to those who are trained with the system with human feedback. %
We carry out t-test between different groups. %
Most of the P-values range from 0.003 to 0.015, also indicates a big or enormous difference between these groups. %
The mean differences in improvement rate between learner trained with \textit{PTeacher} and human feedback system is less than 2.5\%. 
The P results of the t-test between them are 0.29 and 0.42 for learners with high and low proficiency, indicating a minimal difference. %
The result confirms that the effectiveness of exaggerated audiovisual feedback and personalization mechanism. %

Participant interview is conducted during the experiments mentioned above. %
The feeling of participants about the system is asked and recorded. %
More than 85\% of the learners and  80\% of the teachers praise the audio exaggeration. A native speaker says:
\emph{``As an English teacher, I think exaggerated audio feedback is useful. It reminds me of how I teach a learner to say a word right in class."} %
A learner with high proficiency tells us:
\emph{``The audio feedback reminds me of my high school English teacher. She would rectify our mispronunciations when we made mistakes. This method is quite useful for me."} %
A learner with low proficiency also praises the mechanism:
\emph{``I am so happy with this audio exaggeration. It points out the mistake that I can hardly notice."} %

More than 70\% of the learners mention visual exaggeration. %
A learner with low proficiency says:
\emph{``Exaggerated visual feedback is helpful to me. I can imitate the animation and rectify my pronunciation. I do hope that you can try to combine virtual reality technology with visual feedback. That will be even more helpful to me."} %

About 75\% of the learners say that the interactive course can raise their learning interest.%
~A learner with low proficiency says \emph{``The interactive course is fascinating. %
I am willing to spend more time rectify my pronunciation with it. Please design more courses in the future."} %

It is also worth noticing that about 35\% of the learners mention that our system helps deal with educational inequity. %
A learner with low proficiency tells us \emph{``Before I went to college, I had been living in a small town, where the education resource is limited. %
I didn't have any chance to get a foreign teacher to teach me. %
Thus, my English pronunciation training is relatively inadequate. %
With the \textit{PTeacher}, I can learn the pronunciation effectively anytime, anyplace. %
I can't be more willful to introduce it to the children in my town."} %

\section{Discussion}
\subsection{Contributions to HCI}
Based on these studies and the previous learning theories on exaggerated feedback as discussed in Section~\ref{sec:2}, we focus on constructing a participatory exaggerated computer-aided pronunciation training system, \textit{PTeacher}. It emphasizes enlarging the teaching effectiveness from a user perception perspective with two critical aspects: 1) we study how to identify suitable exaggerated feedback to learners with different demands or behaviors in language learning (i.e. different pronunciation proficiencies), and 2) how to define the best set of feedback. Our idea of involving exaggerated feedback and our system of determining the best set of exaggeration parameters would be an important finding to the general area of computer-aided language learning~\cite{garcia2018self, neri2008effectiveness, robertson2018designing, hailpern2009creating} and exaggerated feedback~\cite{granqvist2018exaggeration} systems in HCI.

We further deepen the discussion to general educational systems. We point out that such an idea can be easily migrated to not only other language learning systems but also any imitation-learning system where certain degrees of exaggeration in educational feedback would enhance learners’ perceptibility. For example, under dancing and piano teaching scenarios, we can exaggerate the teaching effect such as increase the music’s expression for piano or the expressiveness of dance animation. The method for how to exaggerate the audio or visual modality can give us much inspiration. However, maybe it is not applicable to logical induction systems such as Math teaching.

\subsection{Limitations}
\noindent\textbf{The Limitations of MDD.}
Though the accuracy of existing \textit{MDD} is already very high according to~\cite{li2016mispronunciation,li2015integrating,li2015rating}, the feedback is still likely to contain two kinds of errors~\cite{badin2010visual}: false accepts (FA) which means that the pronunciation is accepted although it is actually incorrect; and false rejects (FR) means that the pronunciation is rejected although actually correct. As a result, the \textit{MDD} may miss detect mistakenly spoken phonemes or mark correct ones as incorrect, which will affect our system.

There are two traditional mechanisms of pronunciation feedback. One is that the system directly tells the user which sound is mispronounced based on the \textit{MDD} results. The other is to provide an \textit{MDD} score. PTeacher will not directly tell the user which sound is mispronounced but presents the mispronounced phoneme by exaggeration.
As a result, anytime the \textit{MDD} makes a mistake of FR, the learners will receive an exaggerated feedback, or the system will not exaggerate the target mistakenly pronounced word if it is a FA. So the negative impact is mostly that the learner cannot receive the exaggeration in the system’s feedback.

\noindent\textbf{The Limitations of PTeacher.}
The levels of exaggeration are defined mainly in a hand-crafted manner by consulting with English pronunciation education experts, animation designers then leveraging the theory of speech articulators~\cite{browman1992articulatory, kelly2006teach, kozhevnikov1967speech, chang1993animation, thomas1995illusion}. As a result, on the one hand, the designing procedures for exaggerations are time-consuming. On the other hand, the manually defined exaggeration levels can only be limited to a relatively small scale. Moreover, they are not continuously changeable. In the future, we can derive automatic procedures with the help of deep learning technologies~\cite{zhou2019talking,Yang:2020:MakeItTalk} from the collected data.
As for the methodology, only 2 levels of proficiency are defined at phone-level. The personalized feedback can be improved by involving more detailed modelling on the feedback levels of proficiency.

\section{Conclusion}
In this paper, we present \textit{PTeacher}, a pronunciation training system with personalized exaggerated audio-visual corrective feedback and practical training courses. 
Importantly, we uncover how to define the appropriate degree of exaggerations through extensive user-participated experiments.
The optimum set of exaggerations can thus be identified for each individual learner. 
Moreover, interactive training courses are proven to be efficient in improving users' English proficiency.

\begin{acks}
This work is supported by the National Key R\&D Program of China under Grant No. 2020AAA0108600 and No. 2019YFB1405700, the state key program of the National Natural Science Foundation of China (NSFC) (No.61831022) and Tiangong Institute for Intelligent Computing, Tsinghua University. We would like to thank Professor Helen Meng, Associate Research Fellow Chun Yu and Jingbei Li.
\end{acks}

\balance

\bibliographystyle{ACM-Reference-Format}
\bibliography{acmart}

\end{document}